\documentclass[12pt,preprint]{aastex}
\usepackage{bbm}
\usepackage{mathrsfs}
\usepackage{amssymb,amsmath,float}
\usepackage{rotating}
\usepackage{color}

\newcommand\cbr{{\bf \mathfrak{r}}}

\newcommand\cE{\mathcal{E}}

\newcommand\al{\alpha}

\newcommand\de{\delta}
\newcommand\ep{\epsilon}

\renewcommand\th{\theta}

\newcommand\rh{\rho}

\newcommand\ta{\tau}

\newcommand\vp{\varphi}

\newcommand\om{\omega}

\newcommand\De{\Delta}

\newcommand\Om{\Omega}

\newcommand\bna{\bold{\nabla}}
\newcommand\<{\langle}
\renewcommand\>{\rangle}

\newcommand\ie{\emph{i.e.}}
\newcommand\eg{\emph{e.g.}}

\newcommand\beq{\begin{equation}}
\newcommand\eeq{\end{equation}}
\newcommand\bea{\begin{eqnarray}}
\newcommand\eea{\end{eqnarray}}
\newcommand\bal{\begin{align}}
\newcommand\eal{\end{align}}

\newcommand\X{\times}
\newcommand\fr{\frac}

\newcommand\half{{\textstyle \frac{1}{2}}}
\newcommand\ap{\approx}
\newcommand\cd{\cdot}

\newcommand\bj{\bold{j}}
\newcommand\bk{\bold{k}}

\newcommand\bn{\bold{n}}

\newcommand\br{\bold{r}}
\newcommand\bs{\bold{s}}

\newcommand\bzero{\bold{0}}

\renewcommand\bal{\mbox{\boldmath$\alpha$}}

\begin{document}

\title{Large scale density perturbations from a uniform distribution by wave transport}

\author{Richard Lieu$^1$}

\affil{$^1$Department of Physics, University of Alabama,
Huntsville, AL 35899\\}

\begin{abstract}
It has long been known that a uniform distribution of matter cannot produce a Poisson distribution of density fluctuations on very large scales $1/k > ct$ by the motion of discrete particles over timescale $t$.
The constraint is part of what is sometimes referred to as the Zel'dovich bound.  We investigate in this paper the transport of energy by the propagation of waves emanating {\it incoherently} from a regular and infinite lattice of oscillators, each having the same finite amount of energy reserve initially.  The model we employ does not involve the expansion of the Universe; indeed there is no need to do so, because although the scales of interest are all deeply sub-horizon the size of regions over which perturbations are evaluated do far exceed $ct$, where $t$ is the time elapsed since a uniform array of oscillators started to emit energy by radiation (it is assumed that $t$ greatly exceeds the duration of emission).  We find that to lowest order, when only wave fields $\propto 1/r$ are included, there is exact compensation between the energy loss of the oscillators and the energy emitted into space, which means $P(0)=0$ for the power spectrum of density fluctuations on the largest scales.  This is consistent with the Zel'dovich bound; it proves that the model employed is causal, has finite support, and energy is strictly conserved.  To the next order when near fields $\propto r^{-2}$ are included, however, $P(0)$ settles at late times to a positive value that depends only on time, as $t^{-2}$ (the same applies to an excess (non-conserving) energy term).  We further observe that the behavior is peculiar to near fields.  Even though this effect may give the impression of superluminal energy transport, there is no violation of causality because the two-point function vanishes completely for $r>t$ if the emission of each oscillator is sharply truncated beyond some duration.
The result calls to question any need of enlisting cosmic inflation to seed large scale density perturbations in the early Universe.
\end{abstract}


\section{Introduction}

The cosmic inflation theory pioneered by \cite{gut81}, \cite{lin82}, and \cite{alb82} provide a natural explanation of the flatness and isotropy of the observable universe and the observed power spectrum of density perturbations, though attaining the right amplitude requires fine tuning, see \cite{pad93}, \cite{vac00}, \cite{kaw03}, and \cite{yam03}.  Often seen as the strongest argument in its favor is the belief that
without it no {\it a priori} causal mechanism could generate these large-scale density perturbations.  This belief is founded upon the Zel'dovich limit as outlined in \cite{zel65} and \cite{zel83}, {\it viz.} starting from an initially
homogeneous state, random displacements of particles in a manner
consistent with causality (which includes the imposition of the speed of light as the maximum speed of particles and quanta) and the conservation laws cannot lead to a power spectrum $P(k)$ that goes to zero as $k \to 0$ more slowly than $k^4$; more precisely if $P(k) \propto k^n$ then $n \geq 4$.
Observations on the other hand have revealed from the
moment reliable data became available, \cite{spe03}, that $n \approx 1$.  (A special feature is that when $n=1$, the fractional density fluctuation in a region of given size $R$ is finite in both limits $R\to 0$ and $R\to\infty$.)  A detailed discussion
of the physical reasons for the Zel'dovich limit has been given by \cite{gab04}, who suggested (without proof) that quantum effects might invalidate the result.  The limit has also been shown to apply in the context of general relativity \cite{abb86}.

The purpose of this paper is to show by means of a simple example
that if the energy scales permit the quanta responsible for the
perturbations to exhibit predominantly its wave rather than particle
nature -- a rather natural criterion for the early universe -- then
the Zel'dovich limit would indeed be invalidated. This opens the
possibility of finding a causal, non-inflationary mechanism for the generation of the density perturbations.

We begin by reviewing briefly the derivation of the Zel'dovich
limit. Suppose we have a universe filled with massive particles
whose density distribution is homogeneous on all scales
significantly above the mean particle separation, $l$ say, and that some causal process induces random changes in the particle
positions, consistent with the conservation laws of energy and
momentum.  How will this affect the power spectrum $P(k)$ of large-scale density perturbations?
Let us consider time scales much less than the Hubble
time (though large compared to $l/c$), so that we may ignore the
expansion of the universe (the argument also applies in a Friedmann--Robertson--Walker universe provided we make an appropriate choice of gauge, \eg, the synchronous gauge).

Assuming that the particle masses and positions are $m_j$ and $\br_j$, and the Fourier amplitude of the density contrast is $\de_{\bk}$,
such that $P(k)=|\de_{\bk}|^2\ap 0$ for all $k \to 0$, then if the particles are subjected to small displacements $\De\br_l$, the change in $\de_{\bk}$ may be written as
 \beq \De\de_{\bk} \propto \sum_j m_j e^{-i\bk\cd (\br_j + \De\br_j)} = \sum_j m_j e^{-i\bk\cd\br_j}
 \left[-i\bk\cd\De\br_j
 -\fr{1}{2}(\bk\cd\De\br_j)^2+\dots\right].\eeq
Momentum conservation requires that the first term vanishes, so the leading term is the quadratic one.  Consequently we
should expect that, for small $k$, $P(k)=|\De \de_{\bk}|^2 \propto
k^4$. It is conceivable that for some special reason the quadratic
term might also vanish, but in any case we expect $P(k) \propto k^n$
with $n \ge 4$.  In particular, on very large scales $1/k \to \infty$ (or $k \to 0$) one may expect $P(0) = 0$.

However, as pointed out by \cite{gab04}, while this argument
presupposes that the particles are classical as in Newton's corpuscular theory of light, the wave nature of particles and quanta might
modify the picture, because waves do not behave in the same way as
corpuscles, as will be shown explicitly below.



\section{The model}

The model comprises a regular array of classical radiators at the points $\br_i$ of a cubic lattice of spacing $l$ in Minkowski space.  Such a lattice causes the initial power spectrum of density fluctuations to vanish on very large scales (see the explanation after (\ref{balance})), it is therefore a good ansatz for the purpose of testing whether a causal process can seed large scale perturbations capable of overcoming the Zel'dovich bound.  As a result, the energy density averaged over large regions does not fluctuate before the emission of radiation.  We then let each radiator emit scalar field radiation of a given frequency within a limited time period $-\ta \lesssim t \lesssim \ta$, with radiation {\it carrier wave} frequency $\om$ satisfying $\om\ta\gg 1$, and the phase of the radiation emanating from each source is random and uncorrelated with other sources.  The radiators will emit until all their energy is dissipated into radiation.  We shall evaluate energy density perturbations over volumes $r \gg \ta$, and at times $t$ much smaller than the age of the Universe {\it and} $r>t\gg\tau$ (the former condition is to avoid the effects of space expansion).

Such a scenario of matter decaying into radiation is similar to the one employed by slow roll inflation theory for the reheating phase, when the inflaton field undergoes oscillations around the bottom of the slow-roll potential and decays into photons and other relativistic particles.  The only difference is that in the case of inflation density perturbations on superhorizon scales were already present by the era of reheating, whereas the model presented here does not require any pre-existing perturbations on large scales, but relies only on {\it a priori} causal processes in the dissipation phase, which we shall assume to be short compared to the age of the Universe at the time, \ie~$\ta \ll r_H$ ($c=\hbar=1$ here and henceforth) where $r_H$ is the relevant horizon size, to seed them.  Such an assumption enables us to ignore the expansion of the Universe during dissipation.  The purpose of our model, however, is to demonstrate the possibility of generating perturbations on scales $\gg \ta$ (but $\ll r_H$) in time $\ta$, whilst satisfying the requirement of causality.

For simplicity, we restrict our attention to scalar-field radiation.  It would not be difficult to construct a similar model involving electromagnetic radiation, but the final conclusion will be the same.  It would also be possible to generalize to the scenario of radiators emitting a range of frequencies incoherently.

The energy density $u$ and energy flux density (or momentum density) $\bj$ in the real scalar field $\phi (t,\br)$ are given by \beq u = \tfrac{1}{2} [\dot\phi^2 + (\bna\phi)^2],~\bj=-\dot\phi\bna\phi, \label{uj} \eeq where
\beq \phi (t,\br) = \sum_i \phi_i (t,\br) \label{phii} \eeq is the field at any point in space and time due to the array of radiators, with
\beq \phi_i(t,\br)=\fr{1}{r'_i}f(t                            -r'_i)\cos[\om(t-r'_i) + \al_i],~\br'_i = \br_i - \br ,~r'_i = |\br'_i|, \label{phi1} \eeq where $\al_i$ denotes the incoherent phases.
Thus $u$ and $\bj$ are double sums, \beq u=\sum_{i,k}u_{ik},~\bj=\sum_{i,k} \bj_{ik}, \label{uj} \eeq in which \beq u_{ik} = \tfrac{1}{2} (\dot\phi_i\dot\phi_k + \bna\phi_i\cdot\bna\phi_k);~\bj_{ik} = -\dot\phi_i\bna\phi_k. \label{ujik} \eeq
The function $f(t)$ satisfies \beq \left|\fr{\dot f}{f}\right| \ap \fr{1}{\ta} \ll \om, \label{slow} \eeq  \ie~it is assumed to be slowly varying.

Thus $f(t)$ is ideally an exponential function in $|t|$ with time constant $\ta$ such that the total energy emitted over all times is finite.  The other crucial feature of $f(t)$ is that for times $|t| \gg \ta$ the amount of emission is sufficiently negligible, \ie~each
oscillator has finite support.
The outgoing wave at a time $t \gg \ta$ will comprise a spherical shell of radius $\ap t$ and thickness $\ap \ta$.  It is thin in the sense that $\ta \ll t$ but thick in comparison to the wavelength $2\pi/\om$.  Owing these inequalities, the derivative of $\phi_i$ is dominated by the oscillatory factor, {\it viz.} \beq \dot\phi_i=
 -\fr{\om f(t-r'_i)}{r'_i}\sin[\om(t-r'_i) + \al_i].
 \label{phidoti} \eeq
where we ignored $\dot f$ because of (\ref{slow}).  The other derivative of $\phi$ in (2) is \beq \bna\phi_i=\fr{\br'_i}{r'_i}f(t-r'_i)
 \left(\fr{\om}{r'_i}\sin[\om(t-r'_i) + \al_i]
 -\fr{1}{r'_i{}^2}\cos[\om(t-r'_i) + \al_i]\right).
 \label{gradphii}\eeq

Lastly, it should be emphasized that since the scalar field satisfies the Klein-Gordon equation, $u$ and $\bj$ satisfy the continuity equation
 \beq \dot u+\bna\cd\bj=0, \label{continuity} \eeq
which guarantees local energy conservation.

\section{Energy emitted by the radiators}

The total energy emitted by radiator $i$ is given by \beq E_i=
 \int_{-\infty}^{\infty} dt \oint_{{\cal S}_i} a^2d\Om\,\bn\cd\bj(t,\br), \label{Ei} \eeq
where the space integral is over the surface of the small sphere of radius $a$ surrounding $\br_i$, and we extended the time integral to the whole real line since the integrand effectively vanishes outside the interval $[-\ta,\ta]$.  Writing \beq E_i = \sum_{i,j} E_{ij}, \label{Esum} \eeq the diagonal term is \beq E_{ii}=\int_{-\infty}^{\infty} dt \int_{S_i} a^2d\Om\, \bn\cd\bj_{ii}(t,\br). \label{Eii} \eeq  Of course $\bn=\br'_i/r'_i$.  Now when we substitute (\ref{uj}), (\ref{phidoti}) and (\ref{gradphii}), the time integrals will involve only smooth function multiplying the rapidly varying oscillatory ones, so we may replace the latter by time averages, \ie, the $\sin^2$ function is replaced by $\half$ and the product $\sin\cos$ by $0$.  Since there is no angular dependence in the integrand we can trivially perform the angular integral and obtain
\beq E_{ii} = 2\pi\om^2\int_{-\infty}^\infty dt f^2(t). \label{Eii} \eeq

To find all such other contributions $E_{ij}$ with $i\neq j$, note first of all that when calculating $E_i$, the integration over the small sphere around $\br_i$ gets a non-vanishing contribution if the integrand behaves like $1/r'_1{}^2$ for small $r'_i$.  As far as the derivatives of $\phi_k$ for $k\ne i$ are concerned, they are finite and slowly varying.  This means there is no contribution from $\bj_{ik}$, because the integrand only behaves like $1/r'_i$.  The only non-vanishing contribution comes from $\bj_{ki}=-\dot\phi_k\bna\phi_i$, and specifically from inserting the second term in (\ref{gradphii}).  The smooth non-singular factors may be evaluated at $\br_i$. Thus we get
 \beq E_i = \sum_k E_{ik}, \eeq
where for $k\ne i$
 \bea E_{ik}&=&\int_{-\infty}^{\infty} dt \int_{S_i} a^2d\Om\,
 \bn\cd\bj_{ki}(t,\br) \notag\\
 &=&-4\pi\om\int_{-\infty}^{\infty} dt\,
 \fr{f(t-r_{ik})f(t)}{r_{ik}}
 \sin[\om(t-r_{ik}) + \al_k]\cos(\om t + \al_i), \eea
where
 \beq \br_{ik}=\br_i-\br_k. \label{brik} \eeq
Once again averaging the rapidly oscillatory factor, we then find
 \beq E_{ik}=\fr{2\pi\om}{r_{ik}}\sin(\om r_{ik} + \al_i - \al_k)
 \int_{-\infty}^\infty dt f(t-r_{ik})f(t). \eeq
Note that this vanishes for $r_{ik}>2\ta$ because the ranges in which the two factors in the integrand are non-zero no longer overlap.

It is interesting to combine the contributions $E_{ik}$ and $E_{ki}$ representing the mutual effect of one radiator on another.  Clearly the only effect of interchanging $i$ and $k$ is to interchange the two random phases.  Hence we find
 \beq E_{ik}+E_{ki}=\fr{4\pi\om}{r_{ik}}
 \sin(\om r_{ik}) \cos(\al_i - \al_k)
 \int_{-\infty}^\infty dt f(t-r_{ik})f(t).
 \label{radsik}\eeq

\section{Energy in the field}

We proceed to calculate the energy of the radiation in a large volume at late times $r_H\gg t \gg \ta$, {\it viz.} \beq \cE =\int d^3\br\,u(t,\br), \label{fE} \eeq where $u$ is the energy density as given by (\ref{uj}) and (\ref{ujik}).  Strictly speaking the volume integral has to exclude those small spherical cavities of radius $a$ centered on the radiators, although we may assume the contribution of such cavities are sufficiently small that they make no appreciable difference to $\cE$ if we include the field energy in them, as is done here.


The diagonal terms of $\cE$ are \beq \cE_{ii} =\int d^3\br\,u_{ii} (t,\br).  \eeq
Moreover, we see that the $\dot\phi^2$ and $(\bna\phi)^2$ terms in (\ref{gradphii}) are equal, and so
 \beq \cE_{ii} =\int d^3\br
 \fr{\om^2 f^2(t-r'_i)}{r'_i{}^2}\sin^2[\om(t-r'_i)+\al_i] + \int d^3\br \fr{f^2(t-r'_i)}{r'_i{}^4}\cos^2[\om(t-r'_i)+ \al_i]= \cE_{ii}^{(0)} + \cE_{ii}^{(1)}
 \label{cEii}
 \eeq
We can then change variable from $\br$ to $\br'_i$, and perform the angular integration, and also replace $\sin^2$ and $\cos^2$ terms by $\half$.  Note also that in arriving at (\ref{cEii}) the average of $\sin\cos$ was equated to zero.
The first term then has exactly the same expression as before, {\it viz.}~(\ref{Eii}) for the energy of the radiators
 \beq \cE_{ii}^{(0)} =E_{ii}. \eeq
These contributions contain the non-random (or non-interference) part of the energy in the emitters and the wave field, and their exact balance is the direct consequence of energy conservation.  Even if the averaging over these rapidly oscillating sinusoidal functions is not done (because they do introduce some extremely small but finite errors to the calculation) the reader can verify that the $\sin^2$, $\cos^2$, and $\sin\cos$ terms will still exactly cancel the corresponding (likewise unaveraged) terms in $E_{ii}$ when computing the total energy of oscillator and field. Moreover, it can be shown, by pursuing more of the same type of calculations, that even when the slow variation of $f$ is taken into account in the evaluation of $\dot\phi$ and $\bna\phi$, that this exact balance still holds.  The last term of (\ref{cEii}), $\cE_{ii}^{(1)}$, appears to upset the balance (in the sense that energy is created from nothing), but note that this term is smaller than the preceding term by the factor $\ap \om^2 t^2 \gg 1$; indeed at late times $t \to\infty$ it vanishes altogether.  Such a term is analogous to the flux of outflowing electromagnetic energy due to the near fields of a small oscillating dipole, the amplitudes of which decay with distance as $r^{-2}$, which equals $t^{-2}$ because at time $t \gg\ta$ the emitted wavefront is at distance $r$.  In any case, this $\cE_{ii}^{(1)}$ term does not lead to any random density perturbations.


We now evaluate the interference terms $\cE_{ik}$ in the expression (\ref{fE})  for $\cE$, {\it viz.}~\beq \cE_{ik} =\int d^3\br\,u_{ik}(t,\br). \label{cEik} \eeq  Since we are interested in late times, the factor $f(t-r'_i)$ will ensure that $r'_i$ cannot be small.  Hence the second term in (\ref{gradphii}) is much smaller than the first.  In this case the two contributions $\cE_{ik}$ and $\cE_{ki}$ are identical, and we shall consider them together as we form the lowest order estimate of $\cE_{ik}$:
 \bea 2\cE_{ik}^{(0)} &=&
 \om^2\int d^3\br\,\fr{f(t-r'_i)f(t-r'_k)}{r'_i r'_k}
 \left(1+\fr{\br'_i\cd\br'_k}{r'_ir'_k}\right)\notag\\
 &&\X\,\sin[\om(t-r'_i) + \al_i]\sin[\om(t-r'_k) + \al_k]\notag\\
 \label{E1ik0}\eea
It will be shown in Appendix A that (\ref{E1ik0}) reduces to
 \bea 2\cE_{ik}^{(0)} &=&\fr{2\pi\om}{r_{ik}}
 \int_0^\infty dr'\, f(t-r')f(t-r'-r_{ik})\notag\\
 &&\X\,\big[\sin(\om r_{ik} + \al_i - \al_k)
 +\sin(\om r_{ik} - \al_i + \al_k)\big]\notag\\
 &=&\fr{4\pi\om\sin(\om r_{ik})}{r_{ik}} \cos (\al_i - \al_k)
 \int_{-\infty}^\infty ds\,f(s)f(s-r_{ik})
 \label{low}
 \eea
which is {\it identical} to the expression (\ref{radsik}) found above for the corresponding contribution to the output of radiators $i$ and $k$.  It indicates that to lowest order there is exact compensation of energy between the emitted radiation $E_{ik}$ and field $\cE_{ik}^{(0)}$.  If the former comprises a regular lattice of oscillators having the same amount of energy each to start with, the energy within volumes of size $r\gg t$ will, at late times $t\gg\ta$, still not exhibit fluctuations.  To this order of approximation, therefore, the Zel'dovich bound is upheld in the same way with waves as it was with particles.

Another interesting analysis is the following.  From (\ref{A8}) of Appendix A one finds $s=t-r'$, where, from the words below (\ref{A6}) and from (\ref{phi1}), $r'$ either equals $r'_i = |\br_i - \br|$ or $r'_i - r_{ik}$ .  By putting the observer at $\br =0$, one then sees that the product of the two $f$ functions in (\ref{low}) ensures $r_i$ and $r_k$ are both within $\ap \pm \tau$ from $t$, and $r_{ik} \ap \ta$.  Thus, when one sums $\cE_{ik}$ over all $i\neq k$ to form the off-diagonal contribution to the total field energy at $\br = 0$, one is effectively performing an integral over the emission from all oscillators within a thin spherical shell of radius $t$ and thickness $\ta \ll t$ (any large surface of radius $r \gg t$ has no contribution because the product of the two $f$ functions is totally negligible there).  The fact that this result cancels (\ref{radsik}) means, to order $\cE_{ik}^{(0)}$, the flux crossing the small surfaces surrounding the oscillators is exactly compensated by the flux crossing the large surface of radius $t$, and no net perturbation due to wave interference exists to this order.  The same conclusion applies to another $\br$ because the model has no preferred spatial origin.


Next we include the last term of (\ref{gradphii}) to calculate the next order correction term for $\cE_{ik}$, \ie~
\bea 2\cE_{ik}^{(1)} &=& -\om\int d^3 \br \fr{f(t-r'_i)f(t-r'_k)}{r'_i r^{'2}_k}\fr{\br'_i\cd\br'_k}{r'_i r'_k}\notag\\
&&\X\,\sin[\om(t-r'_i)+\al_i]\cos[\om(t-r'_k)+\al_k]
 \label{E1ik1}\eea
Unlike the lowest order integral (\ref{low}), this higher order integral is {\it not} compensated by the corresponding radiator energy term, \ie~
$\cE_{ik}^{(0)} + \cE_{ik}^{(1)} \neq E_{ik}$.  Evaluation of $\cE_{ik}^{(1)}$ follows the same methodology as $\cE_{ik}^{(0)}$, \ie~the steps are similar to those in Appendix A.  The result is
\beq
2\cE_{ik}^{(1)} = \fr{2\pi\sin(\om r_{ik})}{r_{ik}} \sin (\al_i - \al_k) \int_{-\infty}^\infty ds\,f(s)f(s-r_{ik}).
\label{high}
\eeq
Repeating the argument of the previous paragraph, therefore, it means the emitted flux leaving the small surface enclosing the oscillators is {\it not} compensated by the flux crossing the large surface at radius $t$, \ie~there {\it are} perturbations due to wave interference to this next order of approximation.
In the next section we work out the characteristics of the ensuing energy density fluctuation, specifically in respect of the power spectrum $P(\bk)$ seeded by the ensemble of radiators, and discuss the physical meaning.


\section{Power spectrum}

To lowest order, the result just obtained  means there is exact `compensation': if we allow for the differences between energy emitted by the oscillators and the energy in the field, we will find that the resulting power spectrum satisfies $P(0)=0$. But even more crucially is that when the higher order terms of $\cE_{ik}$ are included, this becomes untrue.

To find the resulting power spectrum one may Fourier transform the density distribution
 \beq \widetilde{\de u}(\bk) = \fr{1}{\sqrt{V}}
 \int d^3\br\, \fr{\de u(\br)}{\bar u} e^{-i\bk\cd\br} \eeq
where $V$ is the fundamental volume, and then
 \beq P(\bk)=\<|\widetilde{\de u} (\bk)|^2\>. \eeq
Here $\de u(\br)$ is the density (including of course the delta function contributions from the radiators) minus the mean density $\bar u$.  If each radiator has an initial energy $E_0$, then the mean density, initially or finally\footnote{When writing down (\ref{balance}) we ignored the $\cE_{ii}^{(1)}$ term of (\ref{cEii}) since this term does not cause spatial energy variations and vanishes initially, and also finally at late times $t\to\infty$.},
is $nE_0$, where $n=1/l^3$ is the number density of radiators.

Thus in terms of the quantities computed above
 \beq \de u(\br)=u(\br)+\sum_i(E_0-E_i)\de_3(\br_i)-nE_0. \label{balance} \eeq
Since we are interested in $P(\bk)$ for small $k$, the terms in $E_0$ are irrelevant because they only contribute delta functions at the points of the reciprocal lattice ($\bk$ values whose components are multiples of $2\pi/l$).  Of course, there is no contribution at $\bk=\bzero$.  Similarly, the `diagonal' terms in $u$ and $E_i$ also cancel.  We are left with a sum over the off-diagonal pairs.  Thus
 \beq \widetilde{\de u}(\bk)=
 \fr{1}{\bar u} \left[\sum_{i<j}\xi_{ij}(\bk)
 +\sum_{i<j}\eta_{ij} (\bk)\right], \label{du} \eeq
where
 \bea \xi_{ij} (\bk)= 2\int d^3 \br~u_{ij} (t,\br) e^{-i\bk\cdot\br} \label{dv}\eea
and
 \beq \eta_{ij} (\bk)= -E_{ij}e^{-i\bk\cd\br_i}-E_{ji}e^{-i\bk\cd\br_j}. \label{dw} \eeq
As already demonstrated in the previous section, $\cE_{ij} - E_{ij} = \cE_{ij}^{(1)}$ to lowest order.  Thus, when $\bk = 0$, (\ref{du}) reduces to $2\sum_{i<j} \cE_{ij}^{(1)}$.

The remaining tasks are to (i) calculate $\bar u$, and perform the (ii) $\sum_{i<k}$ summation, (iii) modulus squaring, and (iv) ensemble averaging.  On (i), the energy density at any given point is
\beq u (t,\br) =\fr{\om^2}{2}\sum_{j,l}\fr{f(t-r'_j)f(t-r'_l)}
 {r'_j r'_l}\left(1+\fr{\br'_j\cd\br'_l}{r'_j r'_l}\right)
 \sin[\om(t-r'_j)+\al_j]\sin[\om(t-r'_l)+\al_l].
 \label{u} \eeq
The average value of $u$ is found by averaging (\ref{u}) over the random phases.  Clearly, the only terms that will contribute are those with $j=l$.  Since we are interested in large times, many different radiators will contribute to the value of $u$ at a particular point.  (These are of course located in a spherical shell of radius $t$ and thickness $\ap 2\ta$, containing many lattice points.)  Thus it is reasonable to replace the sum by an integral, over $\br_1$ say, or equivalently over $\br'_1$.  Thus $\<u\>$ is independent of position, as it must be:
 \beq \bar u = \<u(t,\br)\>=\fr{\om^2}{2l^3}\int
 \fr{d^3\br'_1}{r'_1{}^2} f^2(t-r'_1)
 =\fr{2\pi\om^2}{l^3} \int_{-\infty}^\infty f^2(s) ds,
 \label{uav} \eeq
Let us choose a specific form for $f$, {\it viz.} a Gaussian, $f(t)=\sqrt{a/\ta}\,e^{-t^2/2\ta^2}$, so that, from (\ref{uav}), $\bar u = \<u\> = 2\pi^{3/2}a\om^2/l^3$.

Now we proceed to undertake steps (ii) to (iv) in one go.  Clearly this can be done by computing the four fold summation $\sum_{j\neq l} \sum_{i \neq k} \cE_{jl}^{(1)} \cE_{ik}^{(1)}$ and multiplying the result by the factor $1/4$.  Note however that the contributions from $j=i$, $l=k$ and $j=k$, $l=i$ are equal, because the summand is symmetric under the interchange of indices.  Thus we can select one of the terms and double the answer.  Again replacing the sums by integrals and using the notation $\br'_1=\br_1-\br$, $\br_1^{''} = \br_1-\cbr$ {\it etc}~ we obtain
\bea P(0)
 &=& \fr{1}{32\pi^3 \om^2 a^2 V}\int d^3 \br~d^3 \cbr \fr{d^3\br_1}{r'_1 r_1^2}
 \fr{d^3\br_2}{r'_2 r_2^2}f(t-r'_1)f(t-r'_2)f(t-r_1^{''})f(t-r_2^{''}) \nonumber\\
 &&\X\;\fr{\br'_1\cd\br'_2}{r'_1 r'_2}
 \fr{\br_1\cd\br_2}{r_1 r_2}
 \sin[\om(r'_1-r_1)]\sin[\om(r'_2-r_2)] \nonumber\\
 &=& \fr{1}{32\pi^3 \om^2 a^2}\int d^3 \bs~I(t,\bs),
 \label{u2xii}\eea
where the sine factors come from averaging over the random phases, and for $t\gg\ta$ (the values of $t$ of interest to us) $I(t,\br) \ap a^2 \cos^2 \om r/(\om^2 t^2 r^2)$ in the range $1/\om \ll r \lesssim \ta$ and $I(t,\br) \ap 0$ elsewhere (it is exponentially small for $r\gg\ta$), as demonstrated in Appendix B.  This leads to the final result \beq P(0) \ap \fr{\ta}{\om^4 t^2}. \label{P0} \eeq  Note that the fundamental volume in (\ref{u2xii}) is ultimately canceled by $\int d^3 \cbr$ because the two-point function $I(t,\bs)$ depends spatially upon only the relative separation $\bs = \br - \cbr$.

It is important to stress, as explained towards the end of Appendix B, that the two point function $I(t,\br) = 0$ identically (\ie~it vanishes completely) for $r> 20\ta$ if the function $f(t)$ governing the emission duration of the oscillators is truncated beyond $|t|=t_c=10\ta$ whilst maintaining inequality (2.6).  Since none of the aforementioned calculations are affected by such a truncation (all the integrals have long ago converged), this means if one observes the radiation field at times $t \gg \ta$ (more precisely $t > 20 \ta$) in accordance with our ansatz, the two-point function will vanish exactly for $r > t$ and causality is strictly enforced\footnote{The value of the critical epoch $t_c$ for truncation is not restricted to $t_c=10\ta$.  For any arbitrary $t_c \gg \ta$ the two-point function will vanish identically for $r > t$ and $t>2t_c$.}.  Yet the interesting point is that by (\ref{P0}) the fluctuation power in the mode of longest wavelength $\bk =0$ is a finite constant at any given $t$.

\section{Discussion and conclusion}

If there were not for the exact compensation between the lowest order terms for the radiator and wave field energy, the power spectrum would have been $P(0) \ap \ta/\om^2$, a genuine constant becomes the thermal Poisson distribution $P(0) \ap 1/T^3$ in the case of black body emitters with $\om \ap T$ and $\ta \gtrsim 1/T$ (where $T$ is the temperature).  But in reality this compensation of energy exists, which clearly demonstrates that the model we employed is causal, has finite support, and obeys energy conservation.  The consequence is that, to this order, the power spectrum $P(0)=0$ exactly.

Nevertheless to the next order, when one is dealing with the near fields, one is left with the much smaller and time dependent result of (\ref{P0})  which is below than the aforementioned `thermal' constant by the factor $\om^2 t^2 \gg 1$.  Despite the smallness, $P(0)$ is finite and positive, corresponding to Poisson density fluctuations, at any given subsequent epoch.  However, if one truncates the emission window function $f(t)$ beyond some critical epoch $|t| = t_c \gg \ta$, then then $P(0)$ is unaffected but at times well after emission $t > 2t_c$ the two-point function $I(t,\br)$ vanishes exactly for $r > t$, \ie~the seeding of perturbations by this process is then strictly causal.  In this respect, there {\it is} a fundamental difference from the scenario of particle (corpuscular) transport -- the original scenario of the Zel'dovich bound -- that resulted in $P(0) = 0$ for all times.
In other words, there seems to be no alternative interpretation other than claiming that, unlike particle transport, wave transport is capable of breaching the Zel'dovich bound.

In fact, our line of reasoning is reinforced by examining the scenario of just {\it one} oscillator -- an elementary scenario that clearly obeys causality .  Specifically, whether it be an emission of scalar field radiation or electric dipole radiation, it is well known that for $t \gg\ta$, the propagating energy resides within a thin shell of radius $r=t$ and thickness $\ap\ta$, and that the energy crossing the entire outer spherical surface of the shell equals a constant (\ie a value that does not depend on $r$ and $t$) for the wave field, but drops as $1/r^2 = 1/t^2$ for the near field.  Since the wave field energy exactly equals the energy emitted by the finitely supported oscillator, there is a residual uncompensated energy component which does not vanish at finite $t$, even if no extra energy is ultimately emitted to spatial infinity.  Thus the anomaly is already present in the one oscillator case.  It should then come as no surprise that we find similar behavior in an array of many oscillators.  The only difference is that one generally attaches no physical significance to the single oscillator scenario, but the multiple oscillator scenario has implication to model development of the early Universe.




The ramifications on standard inflationary cosmology is two-fold.  First, in respect of the Zel'dovich bound there is a fundamental difference between transporting energy by particles and by waves -- the former obeys the bound while the latter does not.  Note that if one must consider the wave transport model as having taken into account the `quantum' nature of particles, this quantum phenomenon in the current context does not extend beyond wave interferometric effects, \ie~there is not yet any inclusion of
the full-blown Copenhagen interpretation involving non-locality and entangled states for the oscillators.   Second, while the model presented here is not intended to be a realistic
description of the early Universe, the important point is that if even $n=0$ (\ie~$P(k)=$ constant independently of $k$) is accessible, there is no \emph{a priori} reason to rule out a
non-inflationary causal mechanism for a power spectrum with $n\ap
1$.  It is quite possible for a more realistic variant of our proposed mechanism to be found, that could seed the large scale structures we observe without enlisting inflation at all.

\appendix
\section{Calculation of the non-diagonal contributions to the field energy}

Our task is to evaluate the integral of (\ref{E1ik0}), {\it viz.}
\bea 2\cE_{ik} &=&\int d^3\br \big(\dot\phi_i\dot\phi_k
 +\bna\phi_i\cd\bna\phi_k\big)\notag\\
 &=& \om^2\int d^3\br\,\fr{f(t-r'_i)f(t-r'_k)}{r'_i r'_k}
 \left(1+\fr{\br'_i\cd\br'_k}{r'_ir'_k}\right)\notag\\
 &&\X\,\sin[\om(t-r'_i) +\al_i]\sin[\om(t-r'_k) + \al_k]\notag\\
 \label{E1ik}\eea
Note that
 \beq \br'_i\cd\br'_k=\half(r'_i{}^2+r'_k{}^2-r_{ik}^2). \eeq

The integrand of (\ref{E1ik}) depends on $\br$ only through the two scalars $r'_i$ and $r'_k$, so we can convert the integral to a two-fold integral over these variables.  The result is
 \bea 2\cE_{ik} &=&\fr{2\pi\om^2}{r_{ik}}
 \int_0^\infty dr'_i \int_{|r'_i-r_{ik}|}^{r'_i+r_{ik}}dr'_k\,
 f(t-r'_i)f(t-r'_k)\notag\\
 &&\X\,\fr{(r'_i+r'_k)^2-r_{ik}^2}{2r'_ir'_k}
 \sin[\om(t-r'_i) + \al_i]\sin[\om(t-r'_k) + \al_k].
 \label{E1ik2} \eea
We still have rapidly oscillating functions of the integration variables, so in most regions, where they multiply smooth functions the contributions will cancel out.  The only places where this is not true are close to the limits of integration.


Consider the easily evaluated integral
 \beq \int_0^\infty dx\,e^{-\ep x}\sin(\om x+\beta)
 =\fr{\om\cos\beta+\ep\sin\beta}{\om^2+\ep^2}. \eeq
If $\om\gg\ep$ it reduces to $\cos\beta/\om$.  More generally for any smooth function $F(x)$ which is slowly varying relative to $\om$, we have
 \beq \int_a^b dx\,F(x)\sin(\om x+\beta)
 \ap F(a)\fr{\cos(\om a+\beta)}{\om} - F(b)\fr{\cos(\om b+\beta)}{\om}. \eeq

We can apply this result to evaluate the $r'_k$ integral above, obtaining contributions from both upper and lower limits.  Now when $r'_i<r_{ik}$, the lower limit is at $r'_k=r_{ik}-r'_i$.  In that case, the factor in the numerator of (\ref{E1ik2}) vanishes.  So there is no contribution from the lower limit.  Note also that the contribution from the upper limit must have the opposite sign.  Thus we obtain
 \bea 2\cE_{ik} &=&\fr{4\pi\om}{r_{ik}}
 \bigg\{\int_0^\infty dr'_i\, f(t-r'_i)f(t-r'_i-r_{ik})\notag\\
 &&\X\,\sin[\om(t-r'_i) + \al_i]\cos[\om(t-r'_i-r_{ik} + \al_k)]
 \notag\\
 &&-\,\int_{r_{ik}}^\infty dr'_i\, f(t-r'_i)f(t-r'_i+r_{ik})
 \notag\\
 &&\X\,\sin[\om(t-r'_i) + \al_i]\cos[\om(t-r'_i+r_{ik}) + \al_k]
 \bigg\}. \label{A6} \eea
To simplify this, let us write $r'=r'_i$ in the first integral and $r'=r'_i-r_{ik}$ in the second, so that both run over the same range.  This yields the slightly simpler expression
 \bea 2\cE_{ik} &=&\fr{4\pi\om}{r_{ik}}
 \int_0^\infty dr'\, f(t-r')f(t-r'-r_{ik})\notag\\
 &&\X\,\big\{\sin[\om(t-r') + \al_i]\cos[\om(t-r'-r_{ik}) + \al_k]
 \notag\\
 &&-\,\sin[\om(t-r'-r_{ik}) + \al_i]\cos[\om(t-r') + \al_k]\big\}.
 \eea
Rewriting the products of sine and cosine functions as sums then yields
 \bea 2\cE_{ik} &=&\fr{2\pi\om}{r_{ik}}
 \int_0^\infty dr'\, f(t-r')f(t-r'-r_{ik})\notag\\
 &&\X\,\big[\sin(\om r_{ik} + \al_i - \al_k)
 +\sin(\om r_{ik} - \al_i + \al_k)\big]\notag\\
 &=&\fr{4\pi\om\sin(\om r_{ik})}{r_{ik}} \cos (\al_i - \al_k)
 \int_{-\infty}^\infty ds\,f(s)f(s-r_{ik}). \label{A8}
 \eea

 \section{Calculation of the integral for $P(0)$}

 The integral to be evaluated here is of the form
 \bea I(t,\br)
 &=& \int \fr{d^3\br_1}{r'_1 r_1^2}
 \fr{d^3\br_2}{r'_2 r_2^2}f(t-r'_1)f(t-r'_2)f(t-r_1)f(t-r_2) \nonumber\\
 &&\X\;\fr{\br'_1\cd\br'_2}{r'_1 r'_2}
 \fr{\br_1\cd\br_2}{r_1 r_2}
 \sin[\om(r'_1-r_1)]\sin[\om(r'_2-r_2)], \label{u2xi}\eea
where $t\gg \ta$, the primed coordinates are defined in (\ref{phi1}), and the sine factors come from averaging over the random phases.

To proceed further, let us introduce polar coordinates $r_1,\th_1,\vp_1$, with the $z$ axis in the direction of $\br$.  Then we can transform from $\th_1$ to $r'_1$, using $rr_1\sin\th_1\,d\th_1=r'_1dr'_1$.  It is in fact simpler to use the variables $v_1=r_1+r'_1, w_1=r_1-r'_1$.  The limits of integration are given by the requirement that $r,r_1,r'_1$ are the edge lengths of a triangle: $r<v_1<\infty, -r<w_1<r$.  We also need to evaluate the scalar product $\br_1\cd\br_2$.  The $z$ component of $\br_1$ is $z_1=(r^2+v_1w_1)/2r$ while the magnitude of its transverse component $\rh_1$ is given by $ \rh_1^2=(v_1^2-r^2)(r^2-w_1^2)/4r^2$.  For $\br'$, we have $z'_1=(r^2-v_1w_1)/2r$, while its transverse component is the same.  Now $\br_1\cd\br_2=z_1z_2+\rh_1\rh_2\cos(\vp_1-\vp_2)$.  Thus when we do the $\vp$ integrals, we get
 \beq \int_0^{2\pi} d\vp_1 \int_0^{2\pi} d\vp_2\,
 (\br_1\cd\br_2)(\br'_1\cd\br'_2) =
 2\pi^2 (2z_1z_2z'_1z'_2+\rh_1^2\rh_2^2). \eeq
Substituting into (\ref{u2xi}), we see that although the expression for $\xi$ is a seemingly rather daunting four-fold integral over $v_1,w_1,v_2,w_2$, it reduces to a sum of two terms, each of which is the square of a two-dimensional integral:
\beq I(r) = \fr{2\pi^2}{r^2}(2A^2 +B^2), \eeq
\bea
 A(t,r) &=& \int_r^\infty dv
 \int_{-r}^r dw\,f(t-r_1)f(t-r'_1)
 \fr{r^4-v^2w^2}{r^2(v^2-w^2)(v+w)}\sin\om w,\nonumber \\
 B(t,r) &=& \int_r^\infty dv
 \int_{-r}^r dw\,f(t-r_1)f(t-r'_1)
 \fr{(v^2-r^2)(r^2-w^2)}{r^2(v^2-w^2)(v+w)}\sin\om w. \eea
Here of course $r_1=(v+w)/2,r'_1=(v-w)/2$.

The presence of a rapidly oscillating factor will clearly make all these integrals rather small.  To be specific, let us choose a particular form for $f$, namely a Gaussian, $f(t)=\sqrt{a/\ta}\,e^{-t^2/2\ta^2}$.
Then we find that the basic double integral to be evaluated has the form
 \beq C=\int_r^\infty dv
 \int_{-r}^r dw\,f(t-r_1)f(t-r'_1)\sin\om w = \fr{a}{\ta}\int_r^\infty dv\,e^{-(v-2t)^2/4\ta^2}
 \int_{-r}^r dw\,e^{-w^2/4\ta^2}\sin\om w, \label{C} \eeq
with similar, though slightly more complicated, expressions for the double integrals of $A$ and $B$.  Clearly the $v$ integral becomes negligibly small for $r\gtrsim 2t$.  On the other hand, if $r\ll 2t$ then the $v$ integral simply gives a constant, $2\sqrt{\pi}\ta$.  If $r\gtrsim \ta$, then the current limits on the $w$ integral are irrelevant, \ie~they could be extended to $\pm\infty$.  In that case the $w$ integral in (\ref{C}) is $\sim e^{-\om^2\ta^2}$, which by our assumptions is very small.  In particular, by truncating $f(t)$ to zero beyond $t=\pm 10\ta$, $C$ will vanish completely, {\it viz.}~$C=0$ identically for $r > 20 \ta$. So the only region where there could be a significant contribution is $r\lesssim \ta$.  For $r\ll\ta$ we have $(4\sqrt{\pi}a/\om t)\cos{\om r}$ as the approximate value of $C$.  Moreover, provided $t \gg 20\ta$, $C=0$ for all $r > t$.

In examining the contributions of the other terms, we may note that the integrand will be very small except when $v$ is close to $2t$.  This means that $v \gg r > w$.  Thus $A$ and $B$ are effectively of the same form as (\ref{C}) but with extra factors in the integrand of $-w^2/(2r^2 t)$ and $(r^2-w^2)/(2r^2 t)$, respectively.  It follows that \beq A \ap \fr{2\sqrt{\pi}a}{\om^3r^2 t}[(\om^2r^2-2)\cos{\om r}-2\om r\sin\om r], \label{A} \eeq while $ B\ap A+C$.  In particular, we see that in the region $1/\om\ll r\lesssim \ta$ where $I(t,\br)$ is appreciable, \beq I(t,\br) = \fr{a^2 \cos^2\om r}{(\om r)^2 t^2}. \label{I} \eeq  If $f(t)$ is truncated beyond $|t| = 10\ta$, then $I(t,\br) = 0$ for $r > 20 \ta$ by the argument of the previous paragraph.


\end{document}